\documentstyle[12pt]{article}

\advance \topmargin by -\headheight
\advance \topmargin by -\headsep
     
\evensidemargin \oddsidemargin
\begin{document}
\newcommand{\sect}[1]{\setcounter{equation}{0}\section{#1}}
\renewcommand{\theequation}{\thesection.\arabic{equation}}

\topmargin -.6in
\def\nonu{\nonumber}
\def\rf#1{(\ref{eq:#1})}
\def\lab#1{\label{eq:#1}} 
\def\br{\begin{eqnarray}}
\def\er{\end{eqnarray}}
\def\be{\begin{equation}}
\def\ee{\end{equation}}
\def\0{\nonumber}
\def\lb{\lbrack}
\def\rb{\rbrack}
\def\({\left(}
\def\){\right)}
\def\v{\vert}
\def\bv{\bigm\vert}
\def\lskip{\vskip\baselineskip\vskip-\parskip\noindent}
\relax
\newcommand{\nit}{\noindent}
\newcommand{\ct}[1]{\cite{#1}}
\newcommand{\bi}[1]{\bibitem{#1}}
\def\a{\alpha}
\def\b{\beta}
\def\ca{{\cal A}}
\def\cm{{\cal M}}
\def\cn{{\cal N}}
\def\cf{{\cal F}}
\def\d{\delta}
\def\D{\Delta}
\def\eps{\epsilon}
\def\g{\gamma}
\def\G{\Gamma}
\def\grad{\nabla}
\def\h{ {1\over 2}  }
\def\hc{\hat{c}}
\def\hd{\hat{d}}
\def\hg{\hat{g}}
\def\hp{ {+{1\over 2}}  }
\def\hm{ {-{1\over 2}}  }
\def\k{\kappa}
\def\l{\lambda}
\def\L{\Lambda}
\def\lg{\langle}
\def\m{\mu}
\def\n{\nu}
\def\o{\over}
\def\om{\omega}
\def\O{\Omega}
\def\p{\phi}
\def\pa{\partial}
\def\pr{\prime}
\def\ra{\rightarrow}
\def\rh{\rho}
\def\rg{\rangle}
\def\s{\sigma}
\def\t{\tau}
\def\th{\theta}
\def\ti{\tilde}
\def\wti{\widetilde}
\def\inte{\int dx }
\def\xb{\bar{x}}
\def\yb{\bar{y}}

\def\tr{\mathop{\rm tr}}
\def\Tr{\mathop{\rm Tr}}
\def\partder#1#2{{\partial #1\over\partial #2}}
\def\ds{{\cal D}_s}
\def\wtwo{{\wti W}_2}
\def\lie{{\cal G}}
\def\alie{{\widehat \lie}}
\def\dlie{{\cal G}^{\ast}}
\def\elie{{\widetilde \lie}}
\def\edlie{{\elie}^{\ast}}
\def\hlie{{\cal H}}
\def\wlie{{\widetilde \lie}}

\def\rlx{\relax\leavevmode}
\def\inbar{\vrule height1.5ex width.4pt depth0pt}
\def\IZ{\rlx\hbox{\sf Z\kern-.4em Z}}
\def\IR{\rlx\hbox{\rm I\kern-.18em R}}
\def\IC{\rlx\hbox{\,$\inbar\kern-.3em{\rm C}$}}
\def\one{\hbox{{1}\kern-.25em\hbox{l}}}

\def\PRL#1#2#3{{\sl Phys. Rev. Lett.} {\bf#1} (#2) #3}
\def\NPB#1#2#3{{\sl Nucl. Phys.} {\bf B#1} (#2) #3}
\def\NPBFS#1#2#3#4{{\sl Nucl. Phys.} {\bf B#2} [FS#1] (#3) #4}
\def\CMP#1#2#3{{\sl Commun. Math. Phys.} {\bf #1} (#2) #3}
\def\PRD#1#2#3{{\sl Phys. Rev.} {\bf D#1} (#2) #3}
\def\PLA#1#2#3{{\sl Phys. Lett.} {\bf #1A} (#2) #3}
\def\PLB#1#2#3{{\sl Phys. Lett.} {\bf #1B} (#2) #3}
\def\JMP#1#2#3{{\sl J. Math. Phys.} {\bf #1} (#2) #3}
\def\PTP#1#2#3{{\sl Prog. Theor. Phys.} {\bf #1} (#2) #3}
\def\SPTP#1#2#3{{\sl Suppl. Prog. Theor. Phys.} {\bf #1} (#2) #3}
\def\AoP#1#2#3{{\sl Ann. of Phys.} {\bf #1} (#2) #3}
\def\PNAS#1#2#3{{\sl Proc. Natl. Acad. Sci. USA} {\bf #1} (#2) #3}
\def\RMP#1#2#3{{\sl Rev. Mod. Phys.} {\bf #1} (#2) #3}
\def\PR#1#2#3{{\sl Phys. Reports} {\bf #1} (#2) #3}
\def\AoM#1#2#3{{\sl Ann. of Math.} {\bf #1} (#2) #3}
\def\UMN#1#2#3{{\sl Usp. Mat. Nauk} {\bf #1} (#2) #3}
\def\FAP#1#2#3{{\sl Funkt. Anal. Prilozheniya} {\bf #1} (#2) #3}
\def\FAaIA#1#2#3{{\sl Functional Analysis and Its Application} {\bf #1} (#2)
#3}
\def\BAMS#1#2#3{{\sl Bull. Am. Math. Soc.} {\bf #1} (#2) #3}
\def\TAMS#1#2#3{{\sl Trans. Am. Math. Soc.} {\bf #1} (#2) #3}
\def\InvM#1#2#3{{\sl Invent. Math.} {\bf #1} (#2) #3}
\def\LMP#1#2#3{{\sl Letters in Math. Phys.} {\bf #1} (#2) #3}
\def\IJMPA#1#2#3{{\sl Int. J. Mod. Phys.} {\bf A#1} (#2) #3}
\def\AdM#1#2#3{{\sl Advances in Math.} {\bf #1} (#2) #3}
\def\RMaP#1#2#3{{\sl Reports on Math. Phys.} {\bf #1} (#2) #3}
\def\IJM#1#2#3{{\sl Ill. J. Math.} {\bf #1} (#2) #3}
\def\APP#1#2#3{{\sl Acta Phys. Polon.} {\bf #1} (#2) #3}
\def\TMP#1#2#3{{\sl Theor. Mat. Phys.} {\bf #1} (#2) #3}
\def\JPA#1#2#3{{\sl J. Physics} {\bf A#1} (#2) #3}
\def\JSM#1#2#3{{\sl J. Soviet Math.} {\bf #1} (#2) #3}
\def\MPLA#1#2#3{{\sl Mod. Phys. Lett.} {\bf A#1} (#2) #3}
\def\JETP#1#2#3{{\sl Sov. Phys. JETP} {\bf #1} (#2) #3}
\def\JETPL#1#2#3{{\sl  Sov. Phys. JETP Lett.} {\bf #1} (#2) #3}
\def\PHSA#1#2#3{{\sl Physica} {\bf A#1} (#2) #3}
\def\PHSD#1#2#3{{\sl Physica} {\bf D#1} (#2) #3}

\newcommand\twomat[4]{\left(\begin{array}{cc}  %%   2x2 matrix 
{#1} & {#2} \\ {#3} & {#4} \end{array} \right)}
\newcommand\twocol[2]{\left(\begin{array}{cc}  %%   2 column
{#1} \\ {#2} \end{array} \right)}
\newcommand\twovec[2]{\left(\begin{array}{cc}  %%   2 vector
{#1} & {#2} \end{array} \right)}

\newcommand\threemat[9]{\left(\begin{array}{ccc}  %%   3x3 matrix 
{#1} & {#2} & {#3}\\ {#4} & {#5} & {#6}\\ {#7} & {#8} & {#9} \end{array} \right)}
\newcommand\threecol[3]{\left(\begin{array}{ccc}  %%   3 column
{#1} \\ {#2} \\ {#3}\end{array} \right)}
\newcommand\threevec[3]{\left(\begin{array}{ccc}  %%   3 vector
{#1} & {#2} & {#3}\end{array} \right)}

\newcommand\fourcol[4]{\left(\begin{array}{cccc}  %%   4 column
{#1} \\ {#2} \\ {#3} \\ {#4} \end{array} \right)}
\newcommand\fourvec[4]{\left(\begin{array}{cccc}  %%   4 vector
{#1} & {#2} & {#3} & {#4} \end{array} \right)}

%%%%%%%%%%%%%%%%%%%%%%%%%%%%%%%%%%%%%%%%%%Title page %%%%%%%%%%%%%%%%%%%%%%%%%%%%%%%%%%%%%%%%%%%
%%%%%%%%%%%%%%%%%%%%%%%%%%%%%%%%%%%%%%%%%%%%%%%%%%%%%%%%%%%%%%%%%%%%%%%%%%%%%%%%%%%%%%%%%%%%%%%%
%%%%%%%%%%%%%%%%%%%%%%%%%%%%%%%%%%%%%%%%%%%%%%%%%%%%%%%%%%%%%%%%%%%%%%%%%%%%%%%%%%%%%%%%%%%%%%%

\begin{titlepage}
\vspace*{-2 cm}
\noindent
\begin{flushright}
%IFT--P.018/97\\ 
%\today \\

\end{flushright}

\vskip 1 cm
\begin{center}
{\Large\bf T-Duality in 2-D Integrable   Models  } \vglue 1  true cm
{ J.F. Gomes},
 { G.M. Sotkov} and { A.H. Zimerman}\\

\vspace{1 cm}

{\footnotesize Instituto de F\'\i sica Te\'orica - IFT/UNESP\\
Rua Pamplona 145\\
01405-900, S\~ao Paulo - SP, Brazil}\\

\vspace{1 cm}

\end{center}

\normalsize
\vskip 0.2cm

\begin{center}
{\large {\bf ABSTRACT}}\\
\end{center}
\noindent
The non-conformal analog  of abelian T-duality transformations relating pairs of
 axial and vector integrable models from the non abelian
affine Toda family is constructed and studied in detail.

\end{titlepage}

\section{Introduction}
Abelian T-duality in $U(1)^{\otimes s}$ invariant 2-D conformal field theories (CFT's) and in 
string theory represents a set of specific
canonical transformations that relate pairs of equivalent models sharing the same spectrum, but 
with different $\s$--model-like Lagrangians \cite{alvarez}, \cite{giveon}. 
The axial and vector gauged $G/H$-WZW models provide a vast variety of examples of such pairs of 
T-dual models \cite{kiritsis}, \cite{tseytlin}.  From the other side,
the integrable perturbations of these $G/H$-WZW models have been identified  with the family 
of the so called Non-Abelian Affine Toda
theories \cite{local}, \cite{dual}, \cite{elek}. 
 An important feature of these integrable models (IM's) is their  $U(1)^{\otimes k}, \;\; k\leq s$  
global  symmetry   and the fact that they admit both topological and/or non-topological soliton solutions carrying 
$U(1)^{\otimes k}$ charges as well \cite{elek}, \cite{multi}. 
 Hence, an interesting problem to be addressed is about the 
T-duality of pairs of axial and vector
IM's within this family.  More precisely: whether the perturbation breaks a part (or all) of the isometries 
(i.e., $U(1)^{\otimes s_{CFT}}  $ to 
$U(1)^{\otimes s_{Im}}, \;\;  s_{Im}\leq s_{CFT}$ ) and whether certain {\it non conformal} analog of the 
abelian T-duality
transformations takes place.  The simplest example of a pair of T-dual IM's with only one isometry 
(i.e., $s_{Im} =1$) have been studied in
detail in our recent paper  \cite{elek},\cite{workshop}.  As one expects the mass spectrum of the solitons is indeed 
invariant under the corresponding
non-critical T-duality, but the $U(1)$-charges of the solitons of the axial model  are mapped into the topological 
charges of the solitons of the
vector IM and vice-versa.  An interesting example of T-self-dual IM's is given by the complex sine-Gordon 
\cite{dual} and the the Fateev's IM's \cite{Fat}.

The present paper is devoted to the investigation of T-duality properties of  family of IM's representing 
 relativistic IM belonging to the same hierarchy as 
the Fordy-Kulish  (multi-component) nonlinear Schroedinger model (NLS)\cite{FK}, \cite{jmp}.  
They can be considered as a specific 
Hamiltonian reduction of the
$A_n^{(1)}$- Homogeneous sine-Gordon models \cite{miramontes}.  Their main property is the large global symmetry group 
$SL(N)\otimes U(1)$, i.e., they
admit $N$-isometries, as in the $SL(N+1)/SL(N)\otimes U(1)$-WZW model.  As a consequence the T-duality 
transformations relating the 
corresponding axial and vector IM's of this family are indeed more involved.

The paper is organized as follows.  Sect. 2 contains a brief summary of the general formalism for 
the construction of the 
effective action of a large class of NA-affine Toda theories.  In Sect. 3 we apply these methods for 
the derivation of the Lagrangians of
axial and vector IM's of reduced homogeneous SG-type.  Sect. 4 is devoted to the symmetries of such 
models while in Sect. 5  we
explicitly construct the corresponding T-duality transformations.

\section{NA affine Toda models as gauged two-loop WZW models}

The basic ingredient in constructing massive Toda models is the decomposition of an affine 
Lie algebra $\lie $  in
terms of graded subspaces defined according to a grading operator $Q$, 
\br
\quad [ Q, \lie_l] = l \lie_l, \quad \quad 
\lie =\oplus \lie_l, \quad \quad [\lie_l, \lie_k ] \subset \lie_{l+k}, \; l,k= 0, \pm 1, \cdots 
\label{2.1}
\er
In particular, the zero grade subspace $\lie_0$ plays an important 
role since it is parametrized by the Toda fields.
The grading operator $Q$ induces the notion of negative ($\lie_<$) and positive ($\lie_>$) grade subalgebras
and henceforth the decomposition of a group element in the Gauss form, 
\br
g=NBM
\label{2.3}
\er
where $N=\exp {(\lie_{<})}$, $B=\exp {(\lie_{0})}$ and $M=\exp {(\lie_{>})}$.

The action of the corresponding affine Toda  models can be derived 
 from the gauged two-loop \footnote{ the $\hat G$-WZW model in the case where $\hat \lie $ is an affine Kac-Moody algebra is called two-loop
 WZW model \cite{Aratyn}} Wess-Zumino-Witten (WZW)
 action \cite{tau}, \cite{elek}, 
\br
S_{G/H}(g,A,\bar{A})&=&S_{WZW}(g) \nonumber \\
&-&\frac{k}{2\pi}\int d^2x Tr\( A(\bar{\partial}gg^{-1}-\epsilon_{+})
+\bar{A}(g^{-1}\partial g-\epsilon_{-})+Ag\bar{A}g^{-1}  \) 
\label{2.4}
\er
where $A = A_- \in \lie_{<}, \; \bar A = \bar A_+ \in \lie_{>}$ and  $\eps_{\pm}$ are 
constant elements of grade $\pm 1$.  The action (\ref{2.4}) is invariant under
\br
 g^{\prime}=\alpha_{-}g\alpha_{+}, \quad 
 A^{\prime}=\alpha_{-}A\alpha_{-}^{-1}
+\alpha_{-}\partial \alpha_{-}^{-1},
\quad  
 \bar{A}^{\prime}=\alpha_{+}^{-1}\bar{A}\alpha_{+}
+\bar{\partial}\alpha_{+}^{-1}\alpha_{+},  
\label{2.5} 
\er
 where $\a_{-}  \in G_{<}, \; \a_{+}  \in G_{>}$.
It therefore follows that  $ S_{G/H}(g,A,\bar{A}) =S_{G/H}(B,A^{\pr},\bar{A}^{\pr}) $.
 
Integrating over the auxiliary fields $A$ and $ \bar A$ in the partition function 
\br
Z = \int DA D\bar A DB e^{-S}
\label{part} 
\er
we find the effective action for an integrable model defined on the group $G_0$,
\br
S_{eff}(B) =  S_{WZW}(B)- 
 {{k\o {2\pi}}} \int Tr \( \eps_+ B  \eps_- B^{-1}\) d^2x   
  \label{2.6}
\er
The corresponding equations of motion have the following compact form \cite{LS}
\br 
\bar \pa (B^{-1} \pa B) + [ {\eps_-}, B^{-1}  {\eps_+} B] =0, \quad  
\quad \pa (\bar \pa B B^{-1} ) - [ {\eps_+}, B {\eps_-} B^{-1}] =0
\label{2.7}
\er
It is straightforward to derive from the eqns.  (\ref{2.7}) the  
 chiral conserved currents  associated to the subalgebra $\lie_0^0
\subset \lie_0$ defined as 
$\lie_0^0 = \{ X \in \lie_0, \;\; {\rm such \;\; that } \;\; [ X, \eps_{\pm} ] =0 \}$,
 i.e.,
\br
J_X = Tr \( X B^{-1} \pa B \), \quad \quad \bar J_X = Tr \( X \bar \pa B B^{-1}\), 
\quad \quad \bar \pa  J_X = \pa \bar J_X =0
\label{2.7a}
\er
The conservation of such currents is consequence of the invariance of the action  (\ref{2.6}) 
under the $G_0^0 \otimes G_0^0$ chiral transformation, 
\br
B^{\pr} = \bar \Omega (\bar z) B \Omega (z)
\label{chiraltransf}
\er
where $\bar \Omega (\bar z), \Omega (z) \in G_0^0$.

The fact that the currents $J_X$ and  $\bar J_X$  in (\ref{2.7a})  are chiral, allows    further 
reduction of the IM (\ref{2.6}) by imposing a set of  subsidiary
constraints,
\br
J_X = Tr \( X B^{-1} \pa B \) =0 , \quad \quad \bar J_X = Tr \( X \bar \pa B B^{-1}\) = 0, \quad X \in \lie_0^0 
\label{2.7b}
\er
which reduces the model defined on the group $G_0$ to the one on coset $G_0/G_0^0$.
 Such constraints are  incorporated into the action by  repeating the   gauged WZW action argument for the subgroup $G_0$.
 For a general non abelian $\lie_0^0$ we  define a second grading structure $Q^{\pr}$ which decomposes 
 $\lie_0^0$ into positive, zero and negative graded subspaces, i.e., 
 $\lie_0^0 =  \lie_0^{0,<} \oplus \lie_0^{0,0} \oplus \lie_0^{0,>} $.
 Following the same principle as in  \cite{tau}, \cite{elek} and \cite{multi} 
we seek for an action  invariant under
\br
 B^{\pr \pr } = \g_0(\bar z,z) \g_-(\bar z,z)B \g_+(\bar z,z)\g_0^{\pr}(\bar z,z), \quad
 \g_0,  \g_0^{\pr} \in G_0^{0,0},\;  \g_{-} \in G_0^{0,<}, \;  \g_{+} \in G_0^{0,>}
 \label{axvec}
 \er 
and choose 
 $\g_0(\bar z,z), \g_0^{\pr}(\bar z,z), \g_-(\bar z,z), \g_+(\bar z,z) \in G_0^0$ such that 
$B^{\pr \pr } = \g_0 \g_- B \g_+ \g_0^{\pr} = g_0^f \in G_0/G_0^0$ since $B$ 
can also be decomposed into the Gauss form according
to the second grading structure $Q^{\pr}$.
Denote $\G_- = \g_0 \g_-$ and $\G_+ = \g_+ \g_0^{\pr} $.
 Then the action  
 \br 
S(B,{A}^{(0)},\bar{A}^{(0)} ) &=&   S_{WZW}(B)- 
 {{k\o {2\pi}}} \int Tr \( \eps_+ B  \eps_- B^{-1}\) d^2x\nonu \\   
  &-&{{k\o {2\pi}}}\int Tr\( \eta  A^{(0)}\bar{\partial}B
B^{-1} + \bar{A}^{(0)}B^{-1}\partial B +
\eta  A^{(0)}B\bar{A}^{(0)}B^{-1} + A^{(0)}_0\bar{A}^{(0)}_0 \)d^2x \nonu \\
\label{aa}
\er
(with $\eta = +1, -1 $ correspond to $\g_0^{\pr} = \g_0$ for axial  or $\g_0^{\pr} = \g_0^{-1}$ for vector 
gaugings\footnote{ Notice that for non abelian $\lie_0^0$ the invariance of the vector action  in
(\ref{aa}) is consequence of the Borel structure of the subgroup  elements $\Gamma_{\pm}$, i.e., we consider the left-right
 coset ${\Gamma_-}\setminus {G}/{\Gamma_+}$.} respectively, 
$A^{(0)} = A^{(0)}_0 + A^{(0)}_{-}$ and $\bar A^{(0)} = \bar A^{(0)}_0 + \bar A^{(0)}_{+}$),
is invariant under $\Gamma_{\pm}$ transformations
\br
B^{\pr} &=& \G_- B \G_+, \quad 
{A_0^{\pr 0}} = A_0^{(0)}  - \eta \g_0^{-1} \pa \g_0, \quad \quad 
{\bar A^{\pr 0}}_0 = \bar A_0^{(0)}  - \g_0^{-1} \bar \pa \g_0, \nonu \\
A^{\pr (0)} &=& \G_- A^{(0)}\G_-^{-1}  - \eta \pa \G_- \G_-^{-1}, \quad \quad 
\bar A^{\pr (0)} = \G_+^{-1}\bar A^{(0)}\G_+  - \G_+^{-1} \bar \pa \G_+ ,
\label{transf}
\er
where $  A_0^{(0)}, \bar A_0^{(0)} \in \lie_0^{0,0}, A^{(0)}_{-} \in \lie_0^{0,<}, \bar A^{(0)}_{+} \in \lie_0^{0,>} $.  
Hence we have
\br
S(B,{A}^{(0)},\bar{A}^{(0)} ) = S(g_0^f,{A^{\pr}}^{(0)},\bar{A^{\pr}}^{(0)} )
\label{ss}
\er

The general construction above  provides
a systematic classification of relativistic integrable models in terms of its algebraic structure, 
i.e. $\{ \lie , Q, \eps_{\pm}, \lie_0^0 \}$.
 For example, within the affine $\lie = \hat {SL}(N+1)$ algebra we have the following families of integrable models:
\begin{enumerate}
\item  $ \lie_0^0 = \emptyset $ characterizes the choices of  
\br
Q&=& (N+1)d + \sum_{l=1}^{N} \l_l\cdot H, \quad 
 \lie_0 = U(1)^N = \{h_1, \cdots , h_N \}, \nonu \\
\eps_{\pm} &=& \mu (\sum_{l=1}^{N} E_{\pm \a_l}^{(0)} + E_{\mp (\a_1 + \cdots + \a_N)}^{(\pm 1)}) \nonu 
\er
which gives rise  to the well known abelian affine Toda model (see for instance \cite{Olive}, \cite{LS}).
\item  
\begin{enumerate}
\item $\lie_0^0 = U(1) = \{ \l_1 \cdot H \}$
\br
Q&=& N d + \sum_{l=2}^{N} \l_l\cdot H, \quad 
\lie_0 = SL(2)\otimes U(1)^{N-1} = \{E_{\pm \a_1}, h_1, \cdots , h_N \}, \nonu \\
\eps_{\pm} &=& \mu (\sum_{l=2}^{N} E_{\pm \a_l}^{(0)} + E_{\mp (\a_2 + \cdots + \a_N)}^{(\pm 1)})\nonu 
\er
corresponds to the simplest non abelian affine Toda model of dyonic type, admiting electrically charged topological solitons 
(see for instance \cite{tau}, \cite{elek}).
\item  $ \lie_0^0 =U(1)\otimes U(1) = \{ \l_1 \cdot H, \l_N \cdot H \} $ 
\br
Q&=& (n-1) d + \sum_{l=2}^{N-1} \l_l\cdot H, \quad 
\eps_{\pm} = \mu (\sum_{l=2}^{N-1} E_{\pm \a_l}^{(0)} + E_{\mp (\a_2 + \cdots + \a_{N-1})}^{(\pm 1)}) \nonu \\
\lie_0 &=& SL(2)\otimes SL(2)\otimes U(1)^{N-2} = \{E_{\pm \a_1}, E_{\pm \a_N}, h_1, \cdots , h_N \}, \nonu 
\er
is of the same class of $U(1)^{\otimes k}$ dyonic type IM's, but now yielding multicharged solitons (\cite{multi}).
\end{enumerate}
\item   $\lie_0^0 =SL(2)\otimes U(1) = \{ E_{\pm \a_1}, \l_1 \cdot H, \l_2 \cdot H \} $
\br
Q&=& (N-1) d + \sum_{l=3}^{N} \l_l\cdot H, \quad 
\eps_{\pm} = \mu (\sum_{l=3}^{N} E_{\pm \a_l}^{(0)} + E_{\mp (\a_3 + \cdots + \a_{N})}^{(\pm 1)}) \nonu \\
\lie_0 &=&  SL(3)\otimes U(1)^{N-2} = \{E_{\pm \a_1}, E_{\pm \a_2},  E_{\pm (\a_1+\a_2)}, h_1, \cdots , h_N \}, \nonu 
\er
and $Q^{\pr} = \l_1 \cdot H$,  such that 
$\lie_0^{0,<} = \{E_{-\a_1} \}, \;\; \lie_0^{0,>} = \{E_{\a_1} \}, \;\; \lie_0^{0,0} = \{\l_1\cdot H, \l_2\cdot H \}$
leads to dyonic models with non abelian global symmetries (see Sect. 6 of \cite{multi}).
\end{enumerate}
%\end{enumerate}

The classical integrability of all  these models follows from their zero curvature (Lax) representation:
\br
\pa \bar {{\cal A}} - \bar \pa {\cal A} - [{\cal A},\bar {{\cal A}}] = 0, \quad \quad {\cal A},\bar {{\cal A}} \in \oplus_{i=0, \pm 1}
\lie_i
\label{zzc}
\er
with 
\br
{\cal A} = -B \eps_-B^{-1}, \quad \bar {{\cal A}} =\eps_+ + \bar \pa B B^{-1}
\label{zz}
\er
where the constraints (\ref{2.7b}) are imposed.  It can be easily verified that substituting (\ref{zz}) into (\ref{zzc}) 
taking into account (\ref{2.7b}), one reproduces the equations of motion  (\ref{2.7}).  Then the existence of  an infinite set (of commuting)
conserved charges $P_m, \;\; m=0,1, \cdots$ is a simple consequence of eqn. (\ref{zzc}), namely, 
\br
P_m (t)= Tr \( T(t) \)^m, \quad \pa_t P_m = 0, \quad 
T(t) =\lim \limits_{L \to \infty } {\cal P} exp \int_{-L}^{L} {\cal A}_x(t,x)dx
\nonu
\er
Hence the above described procedure for derivation of the abelian and NA affine Toda models as gauged $G/H$ two loop WZW models leads to
integrable models by construction.

\section{Homogeneous Gradation and the Lund-Regge Type Models}

An interesting class of integrable models, that generalizes the Lund-Regge model \cite{L-R}, 
 can be constructed from the affine Kac-Moody algebra $\hat \lie = \hat {SL}(N+1)$ endowed with
   homogeneous gradation $Q=d$ and the specific choice of 
 $\eps_{\pm} = \mu \l_N \cdot H^{(\pm 1)}$, 
 where $\l_N$ is the $N^{th}$ fundamental weight of $SL(N+1)$.
The zero grade subalgebra $\lie_0$  corresponds to the finite dimensional Lie 
algebra $\lie_0 = SL(N+1)$ and  $\lie_0^0 = SL(N)\otimes U(1)$.  
Let us parametrize the auxiliary gauge fields as follows
\br
 A_0^{(0)}&=& \sum_{i=1}^{N} a_{i} (\l_i - \l_{i-1}) \cdot H^{(0)}, \quad \quad 
 \bar A_0^{(0)}= \sum_{i=1}^{N} \bar a_{i} (\l_i - \l_{i-1}) \cdot H^{(0)}, \quad \l_0 =0 \nonu \\
 A_{-}^{(0)}&=& \sum_{j=1}^{N-1}\sum_{i=j}^{N-1} a_{i+1,j} E_{-(\a_j +\cdots + \a_i)}^{(0)},  \quad \quad 
 \bar A_+^{(0)}= \sum_{j=1}^{N-1}\sum_{i=j}^{N-1} \bar a_{j,i+1} E_{\a_j +\cdots + \a_i}^{(0)}
\label{aaaa}
\er
where $a_{ij}(x,t), a_i(x,t), \bar a_{ij}(x,t), \bar a_i(x,t)$ are arbitrary functions of space-time variables.  We next consider  two
different gauge fixings of $\lie_0^0$, the vector and the axial,  in order to derive the effective Lagrangians for the pair of T-dual IM's. 

\subsection{Axial Gauging}

%The coset $\lie_0 /\lie_0^0$ is then spanned by 
%$\lie_0 /\lie_0^0 = \{ E_{\pm (\a_N + \a_{N-1} + \cdots +\a_i)}, i=1, 2, \cdots N \} $, such that 
According to the axial gauging (\ref{axvec}), $\eta = 1, \g_0^{\pr} = \g_0$, 
the factor group element $g_0^f \in G_0/G_0^0$ is parametrized as follows
 \br
g_0^{f} =    g_{0, ax}^{f} = n m, \quad n=e^{\sum_{i=1}^{N} \chi_i E_{-(\a_i + \cdots + \a_N)}}, \;\; 
 m=e^{\sum_{i=1}^{N} \psi_i E_{\a_i + \cdots + \a_N}}
\label{g0f}
\er
After a tedious but straightforward calculation we find
\br
     &&  Tr \( A_0^{(0)} \bar A_0^{(0)}  + A^{(0)} g_0^f \bar A^{(0)} g_0^{f -1}  
+ A^{(0)}\bar \pa g_0^f   g_0^{f -1} + \bar A^{(0)}  g_0^{f -1}\pa g_0^f \) \nonu \\
&& =\bar { a}_iM_{ij} { a}_j + \bar { a}_iN_i +  \bar N_i { a}_i  
 + \sum_{j=1}^{N-1} \sum_{i=j}^{N-1}\sum_{k=j}^{N-1}\bar a_{j,i+1}a_{k+1,j} (\d_{i,k} + \psi_{i+1}\chi_{k+1}) \nonu \\
&& - \sum_{j=1}^{N-1} \sum_{i=j}^{N-1}\bar a_{j,i+1} \psi_{i+1}\pa \chi_j-
 \sum_{j=1}^{N-1} \sum_{i=j}^{N-1} a_{i+1,j} \chi_{i+1}\bar \pa \psi_j
\label{tr1}
\er
where we have introduced $M_{ij}$ and $N_j, \bar N_j$ as
\br
M_{i,j} &=& 2(\l_i - \l_{i-1})\cdot (\l_j - \l_{j-1}) + \psi_i \chi_i \d_{i,j},\quad i,j=1, \cdots N, \; \l_0 =0 \nonu
\er
\br
N_j &=& \( \sum_{i=j}^{N-1} a_{ i+1,j} \chi_{i+1} - \pa \chi_j\) \psi_j, \quad 
\bar N_j= \( \sum_{i=j}^{N-1} \bar a_{j, i+1} \psi_{i+1} - \bar \pa \psi_j\) \chi_j, 
\label{mnn1}
\er
In order to derive the effective Lagrangian of the axial model we have to integrate the auxiliary fields $a_1, \bar a_i, a_{j,i+1}$ and $\bar
a_{i+1, j}$.  We shall 
 consider  the particular case  $N=2$, i.e. $\lie = \hat SL(3)$, where 
 the Gaussian matrix integration is quite simple.  Then, in the  parametrization
 (\ref{g0f})
 \br
B &=& e^{\tilde \chi_1 E_{-\a_1}}e^{\tilde \chi_2 E_{-\a_2}+ \tilde \chi_3 E_{-\a_1-\a_2}}
e^{\phi_1 h_1+\phi_2 h_2} e^{\tilde \psi_2 E_{\a_2}+ \tilde \psi_3 E_{\a_1+\a_2}} e^{\tilde \psi_1 E_{\a_1}}\nonu \\
&=&e^{\tilde \chi_1 E_{-\a_1}}e^{{1\o 2}(\l_1 \cdot H R_1 + \l_2 \cdot H R_2)} \( g_{0, ax}^f \)
e^{{1\o 2}(\l_1 \cdot H R_1 + \l_2 \cdot H R_2)}
e^{\tilde \psi_1 E_{\a_1}}\nonu \\
g_{0, ax}^f &=& e^{\chi_1 E_{-\a_1-\a_2}+\chi_2 E_{-\a_2}} e^{\psi_1 E_{\a_1+\a_2}+\psi_2 E_{\a_2}}\nonu \\
&& \phi_1 h_1+\phi_2 h_2 = \l_1 \cdot H R_1 + \l_2 \cdot H R_2
\label{ax}
\er
we have  $M_{ij}, N_j \bar N_j, \; i,j=1,2$ in  the form
\br
M =  \twomat{{4\o 3} + \psi_1 \chi_1}{- {2\o 3}}{-{2\o 3}}{{4\o 3} + \psi_2 \chi_2}, \quad 
\label{matr2}
\er
and 
\br
\bar N = \twovec{-( \bar \pa \psi_1-\bar a_{1,2}\psi_2 )\chi_1 ,}{-(\chi_2\bar \pa \psi_2)}, \quad
 N = \twocol{-(  \pa \chi_1- a_{2,1}\chi_2 )\psi_1}{-(\psi_1\pa \chi_1)}
\label{mat2}
\er
%It therefore follows that
%\br
%N M^{-1}N &=& {{4\Delta }\o {3 D}} a_1 \bar a_1 + 
%2{{\chi_1 a_1}\o {3 D}} \( \chi_2 \psi_1 \bar \pa \psi_2 -2 (1+\psi_2 \chi_2) \bar \pa \psi_1\) \nonu \\
%&+& 2{{\psi_2 \bar a_1}\o {3 D}} \( \psi_2 \chi_1\pa \chi_2 -2(1+ \psi_2 \chi_2)\pa \chi_1\) \nonu \\
%&+& {1\o {3D}} \( -({4} + 3\psi_2 \chi_2)\psi_1 \chi_1 \bar \pa \psi_1 \pa \chi_1 
%- 2\chi_1 \psi_2 \pa \chi_2 \bar \pa \psi_3 \right. \nonu \\
% &-& \left. 2\chi_2 \psi_1 \bar \pa \psi_2
%\pa \chi_1 -  ({4} + 3\psi_1 \chi_1)\psi_2 \chi_2 \bar \pa \psi_2 \pa \chi_2\)
%\label{ieff1}
%\er
%where 
Integrating first over the $a_i$ and $ \bar a_i$  and next on the $a_{12}, \bar a_{21}$  we derive  the  effective 
action of the $SL(3)$ axial
model
\br
S_{ax} = -{{k}\o {2\pi }} \int dz d\bar z & \( {1\o {\Delta}}( {{\bar \pa \psi_2 \pa \chi_2 }} (1 + 
\psi_1\chi_1 + \psi_2 \chi_2 )  + 
{{\bar \pa \psi_1 \pa \chi_1} }(1 + \psi_2 \chi_2 )  \right. \nonu \\
& \left. - {1\o 2}\( \psi_2 \chi_1 {{\bar \pa \psi_1 \pa \chi_2 }}+ \chi_2 \psi_1 {{\bar \pa \psi_2 \pa \chi_1 }}\)) -V \)
\label{6.6}
\er
where $V =  \mu^2({2\o 3} + \psi_1 \chi_1 + \psi_2 \chi_2) $ and 
$\Delta = (1+ \psi_2 \chi_2 )^2 + \psi_1 \chi_1 (1 + {3 \o 4} \psi_2 \chi_2 )$.

\subsection{Vector Gauging}

For the explicit $SL(3)$ case,  the zero grade group element $B$ is written according to the vector gauging ($\eta = -1, \g_0^{\pr} =
\g_0^{-1}$) as
\br
B &=& e^{\tilde \chi_1 E_{-\a_1}}e^{\tilde \chi_2 E_{-\a_2}+ \tilde \chi_3 E_{-\a_1-\a_2}}
e^{\phi_1 h_1 + \phi_2 h_2} e^{\tilde \psi_2 E_{\a_2}+ \tilde \psi_3 E_{\a_1+\a_2}} e^{\tilde \psi_1 E_{\a_1}}\nonu \\
&=&e^{\tilde \chi_1 E_{-\a_1}}e^{{1\o 2}(\l_1 \cdot H u_1 + \l_2 \cdot H u_2)} \( g_{0, vec}^f \)
e^{-{1\o 2}(\l_1 \cdot H u_1 + \l_2 \cdot H u_2)}
e^{\tilde \psi_1 E_{\a_1}}
\label{vec}
\er
where 
$g_{0,vec}^f = e^{-t_2 E_{-\a_2}- t_1 E_{-\a_1-\a_2}} 
e^{\phi_1 h_1 + \phi_2 h_2}e^{t_2 E_{\a_2}+ t_1 E_{\a_1+\a_2}}$.  We next    choose $u_1, u_2$ such that 
\br
\tilde \chi_2 e^{-{1\o 2}u_2} = -t_2, \quad \tilde \psi_2 e^{{1\o 2}u_2} = t_2, \nonu \\
\tilde \chi_3 e^{-{1\o 2}(u_1+u_2)} = -t_1, \quad \tilde \psi_3 e^{{1\o 2}(u_1+u_2)} = t_1
\nonu
\er
Taking into account the parametrization 
 (\ref{aaaa})  for $SL(3)$  we find 
\br
 & & Tr \( A_0^{(0)} \bar {A}_0^{(0)} - A_0^{(0)} g_{0,vec}^f \bar {A}_0^{(0)} g_{0,vec}^{f -1} 
+ \bar   A_0^{(0)}  g_{0,vec}^{f -1}   \pa g_{0,vec}^f                       
-{A}_0^{(0)}  \bar \pa g_{0,vec}^f g_{0,vec}^{f -1} \) \nonu \\
&=& a_1 \bar a_1 \bar \Delta + \bar a_1 (a_{01} t_1t_2  + t_2 \pa t_1) e^{\phi_1 + \phi_2}
+  a_1 (\bar a_{01} t_1t_2  - t_2 \bar \pa t_1) e^{\phi_1 + \phi_2} \nonu \\
&+& a_{01}\bar a_{01} t_1^2 e^{\phi_1 + \phi_2} + a_{02}\bar a_{02} t_2^2 e^{2\phi_2 - \phi_1} 
+ \bar a_{01}(\pa \phi_1 + t_1 \pa t_1 e^{\phi_1 + \phi_2}) + 
\bar a_{02}(\pa \phi_2 - \pa \phi_1  + t_2 \pa t_2 e^{-\phi_1 + 2\phi_2}) \nonu \\
&-& a_{01} (\bar \pa \phi_1 + t_1 \bar \pa t_1  e^{\phi_1 + \phi_2}) - 
a_{02} (\bar \pa \phi_2 - \bar \pa \phi_1  + t_2 \bar \pa t_2  e^{-\phi_1 + 2\phi_2})
\label{acvec}
\er
where $\bar \Delta =   t_2^2 e^{\phi_1 + \phi_2} -e^{2\phi_1 -\phi_2}$. We first take the integral  over $a_1$ and $\bar a_1$ 
in the partition function (\ref{part}) with the action given by (\ref{aa}).    As a result we get 
\br
{\cal L}_{int} &=& \bar { a}_{0i}M_{ij} { a}_{0j} + \bar { a}_{0i}N_i +  \bar N_i { a}_{0i} 
+ {{t_2^2 \pa t_1 \bar \pa t_1 }\o {\bar \Delta}}e^{2(\phi_1 + \phi_2)}
\label{matr}
\er
where
\br
M_{11} = -{{t_1^2 }\o {\bar \Delta}}e^{3\phi_1}, \quad M_{22} = t_2^2 e^{2\phi_2 - \phi_1}, \quad M_{12} = M_{21} = 0
\label{mm} 
\er
 and 
 \br
 N_1 &=& \pa \phi_1 + t_1 \pa t_1 e^{\phi_1 + \phi_2} - {{t_1 t_2^2 \pa t_1 }\o {\bar \Delta}}e^{2(\phi_1 + \phi_2)}, \quad
 N_2 = \pa \phi_2 - \pa \phi_1 + t_2 \pa t_2 e^{-\phi_1 + 2\phi_2}, \nonu \\
 \bar N_1 &=& -\bar \pa \phi_1 - t_1 \bar \pa t_1 e^{\phi_1 + \phi_2} + {{t_1 t_2^2 \bar \pa t_1 }\o {\bar \Delta}}e^{2(\phi_1 + \phi_2)},
  \quad
\bar  N_2 = -\bar \pa \phi_2 + \bar \pa \phi_1 - t_2 \bar \pa t_2 e^{-\phi_1 + 2\phi_2}\nonu \\
\label{nn}
\er
We next integrate the fields $\bar a_{0i}$ and $a_{0i}, \; i=1,2$  in eqn. 
 (\ref{matr}).   Together with the standand form of WZW action $S_{WZW}(g_{0, vec}^f)$ we arrive at the following effective Lagrangian for
 the vector IM, 
 \br
 {\cal L}_{vec} &=& {1\o 2}\sum_{i=1}^{2} \eta_{ij}\pa \phi_i \bar \pa \phi_j +
 {{\pa \phi_1 \bar \pa \phi_1}\o {t_1^2}}e^{-\phi_1-\phi_2} +\bar \pa \phi_1 \pa ln (t_1) +  
 \pa \phi_1 \bar \pa ln (t_1)  -
 {\pa \phi_1 \bar \pa \phi_1}{\( {t_2}\o {t_1}\)^2}e^{-2\phi_1 +\phi_2}  \nonu \\
 &+& {{\bar \pa (\phi_2 -\phi_1)\pa (\phi_2 - \phi_1)}\o {t_2^2}}e^{\phi_1-2\phi_2}
   + \bar \pa (\phi_2 - \phi_1) \pa ln (t_2) +  
 \pa (\phi_2- \phi_1) \bar \pa ln (t_2)  -V 
 \label{svec}
 \er
 where $V = \mu^2({2\o 3}  - t_2^2 e^{-\phi_1 +2 \phi_2} - t_1^2 e^{\phi_1+\phi_2})$ and 
 $\eta_{ij} = 2 \d_{ij}-\d_{i, j-1}-\d_{i, j+1}$. The integrability of the axial (\ref{6.6}) 
 and vector (\ref{svec}) models is a consequence of the Lax representation (\ref{zzc}) and (\ref{zz}) valid for both models.
%%%%%%%%%%%%%%%%%%%%%%%%%%%%%%%%%%%%%%%%%%%%%%%%%%%%%%%%%%%%%%%%%%%%%%%%%%%%%%%%%%%%%%%%%%%%%%%%%%%%%%%%%%%%%%%%

\section{Local and Global Symmetries}

Before imposing the subsidiary constraints (\ref{2.7b}) the model on the {\it group} $G_0$  described by (\ref{2.6}) is 
invariant under {\it chiral} transformation
(\ref{chiraltransf}) generated by $G_0^0 \otimes G_0^0$.
 For the explicit $SL(3)$ case, the  associated Noether currents are given in terms of 
the axial variables defined in
(\ref{ax}) as 
\br
J_{-\a_1} &=& \pa \tilde \psi_1 - \tilde \psi_1^2 \pa \tilde \chi_1 e^{R_1} + \pa \tilde \chi_2 (\tilde \psi_1 \tilde \psi_2 - \tilde
\psi_3)e^{R_2} \nonu \\
& & + (\pa \tilde \chi_3 - \tilde \chi_2 \pa \tilde \chi_1)(\tilde \psi_1 \tilde \psi_2 - \tilde \psi_3) \tilde \psi_1 e^{R_1 +
R_2} + \tilde \psi_1 \pa R_1 ,\nonu \\
J_{\a_1} &=& \pa \tilde \chi_1 e^{R_1} - \tilde \psi_2 (\pa \tilde \chi_3 - \tilde \chi_2 \pa \tilde \chi_1)e^{R_1+R_2},\nonu \\
J_{\l_1 \cdot H} &=& {1\o 3} (2\pa R_1 + \pa R_2) - \tilde \psi_1 \pa \tilde \chi_1 e^{R_1} + (\tilde \psi_1 \tilde \psi_2 - \tilde \psi_3)
(\pa \tilde \chi_3 - \tilde \chi_2 \pa \tilde \chi_1)e^{R_1+R_2},\nonu \\
J_{\l_2 \cdot H} &=& {1\o 3} (\pa R_1 + 2\pa R_2) - \tilde \psi_2 \pa \tilde \chi_2 e^{R_2} - 
\tilde \psi_3 (\pa \tilde \chi_3 - \tilde \chi_2 \pa \tilde \chi_1)e^{R_1+R_2}\nonu \\
\bar J_{\a_1} &=& \bar \pa \tilde \chi_1 - \tilde \chi_1^2 \bar \pa \tilde \psi_1 e^{R_1} 
+ \bar \pa \tilde \psi_2 (\tilde \chi_1 \tilde \chi_2 - \tilde
\chi_3)e^{R_2} \nonu \\
& &+ (\bar \pa \tilde \psi_3 - \tilde \psi_2 \bar \pa \tilde \psi_1)(\tilde \chi_1 \tilde \chi_2 - \tilde \chi_3) \tilde \chi_1 e^{R_1 +
R_2} + \tilde \chi_1 \pa R_1,\nonu \\
\bar J_{-\a_1} &=& \bar \pa \tilde \psi_1 e^{R_1} - \tilde \chi_2 (\bar \pa \tilde \psi_3 - 
\tilde \psi_2 \bar \pa \tilde \psi_1)e^{R_1+R_2},\nonu \\
\bar J_{\l_1 \cdot H} &=& {1\o 3} (2\bar \pa R_1 + \bar \pa R_2) - \tilde \chi_1 \bar \pa \tilde \psi_1 e^{R_1} 
+ (\tilde \chi_1 \tilde \chi_2 - \tilde \chi_3)
(\bar \pa \tilde \psi_3 - \tilde \psi_2 \bar \pa \tilde \psi_1)e^{R_1+R_2},\nonu \\
\bar J_{\l_2 \cdot H} &=& {1\o 3} (\bar \pa R_1 + 2\bar \pa R_2) - \tilde \chi_2 \bar \pa \tilde \psi_2 e^{R_2} - 
\tilde \chi_3 (\bar \pa \tilde \psi_3 - \tilde \psi_2 \bar \pa \tilde \psi_1)e^{R_1+R_2}
\label{noether}
\er
where $\bar \pa J = \pa \bar J = 0$ and $J = J_{\l_1 \cdot H} h_1 + J_{\l_2 \cdot H} h_2 
+ \sum _{\a}  J_{\a} E_{-\a} + J_{-\a} E_{\a}, \;\; \a= \a_1, \a_2, \a_1+ \a_2$.
Apart from those Noether currents (\ref{noether}) notice the existence of {\it topological} currents 
\br
j_{\varphi, \mu} = \eps_{\mu \nu} \pa_{\nu}\varphi, \quad  \varphi = \{ R_i,i=1,2,\;\; \tilde \chi_j, \tilde \psi_j,  j=1,2,3 \}.
\label{topological}
\er
The reduction from the group $G_0$ to the coset $G_0/G_0^0$ implies the vanishing of currents (\ref{noether}), which defines the unphysical
nonlocal fields $R_i$ in terms of $\psi_i, \chi_i$:
\br
 \pa R_1 &=& {{\psi_1 \pa \chi_1} \o {\Delta}}( 1+{3\o 2}\psi_2 \chi_2)  - {{\psi_2 \pa \chi_2 }\o {\Delta}}( \Delta_2 + 
 {{3\o 2}} \psi_1 \chi_1), \nonu \\
 \pa R_2 &=& {{\psi_1 \pa \chi_1  }\o {\Delta}} + {{\psi_2 \pa \chi_2 }\o {\Delta}}( 2 \Delta_2 + {{3\o 2}} \psi_1 \chi_1), \nonu \\
\bar  \pa R_1 &=&  
{{\chi_1 \bar \pa \psi_1 } \o {\Delta}} ( 1+{3\o 2}\psi_2 \chi_2)  - {{\chi_2 \bar \pa \psi_2 }\o {\Delta}}( \Delta_2 + 
{{3\o 2}} \psi_1 \chi_1),\nonu \\
 \bar \pa R_2 &=&
{{\chi_1 \bar \pa \psi_1  }\o {\Delta}} + {{\chi_2 \bar \pa \psi_2 }\o {\Delta}}( 2 \Delta_2 + {{3\o 2}} \psi_1 \chi_1)
\label{rr}
\er 
where $\Delta = (1+ \psi_2 \chi_2 )^2 + \psi_1 \chi_1 (1+ {3\o 4} \psi_2 \chi_2 ), \quad \Delta_2 = 1+ \psi_2 \chi_2$ and 
\br
 &\tilde \chi_1 = \chi_3 e^{-{1\o 2}R_1},  \quad \quad & \tilde \psi_1 = \psi_3 e^{-{1\o 2}R_1}, \nonu \\
 &\tilde \chi_2 = \chi_2 e^{-{1\o 2}R_2},  \quad \quad & \tilde \psi_2 = \psi_2 e^{-{1\o 2} R_2}, \nonu \\
 &\tilde \chi_3 = \chi_1 e^{-{1\o 2}(R_1 + R_2)},  \quad & \tilde \psi_3 = \psi_1 e^{-{1\o 2}(R_1 + R_2)}. 
 \label{newvaria}
 \er
In addition we find 
 \br
  \pa \tilde \chi_1 &=&  {{\psi_2}\o {\Delta}}\( \pa \chi_1 \Delta_2 -{1\o 2}\chi_1  \psi_2  \pa \chi_2 \)e^{-{1\o 2}R_1}, \nonu \\
 \pa \tilde \psi_1  &=&  
 {{\psi_1}\o {\Delta}} \( \pa \chi_2 (1+ \psi_1 \chi_1 + \psi_2 \chi_2) - {1\o 2} \chi_2 \psi_1 \pa \chi_1\) e^{-{1\o 2}R_1} , \nonu \\
 \bar \pa \tilde \psi_1  &=& {{\chi_2}\o {\Delta}}\( \bar \pa \psi_1 \Delta_2 -{1\o 2} \psi_1\chi_2  \bar \pa \psi_2\)e^{-{1\o 2}R_1}, \nonu \\ 
\bar \pa \tilde \chi_1  &=& {{\chi_1}\o {\Delta}} \( \bar \pa \psi_2 (1+ \psi_1\chi_1 + \psi_2\chi_2) - 
{1\o 2} \chi_1\psi_2 \bar \pa \psi_1 \)e^{-{1\o 2}R_1} . 
  \label{psichi}
  \er
Using the equations of motion derived from (\ref{6.6}), we prove the following conservation laws\footnote{Notice that (\ref{psichi})
 denotes non local fields $R_1, R_2, \tilde \psi_1,\tilde \chi_1$  in terms of the physical fields $\psi_1, \psi_2, \chi_1$ and $\chi_2$ and
 hence conservation of (\ref{noether1}) is non trivial}
 \br
\bar \pa j = \pa \bar j, \quad j=  j_{\tilde \psi_1}, \quad j_{\tilde \chi_1}, \quad j = j_{R_i}, i=1,2, 
\label{jbarj}
\er
where $j = {1\o 2}( j_0 +j_1), \quad \bar j = {1\o 2}( j_0 -j_1)$, and 
\br
j_{R_i, \mu} = \eps_{\mu \nu} \pa_{\nu}R_i, \quad i=1,2, \quad 
j_{\tilde \psi_1, \mu} = \eps_{\mu \nu} \pa_{\nu}\tilde \psi_1, \quad 
 j_{\tilde \chi_1, \mu} =\eps_{\mu \nu} \pa_{\nu}\tilde \chi_1, 
 \label{noether1}
 \er
 Under the reduction (\ref{2.7b}), the topological currents (\ref{topological}) in the group $G_0$ become Noether currents 
 (\ref{noether1}) in 
 the coset $G_0/G_0^0$ and their   conservation   is consequence of the  
 invariance of action (\ref{6.6}) under the following nonlocal global transformations 
 \br
 \d \psi_1 &=& {1\o 2} (-\eps_1 -\eps_2 + \bar \eps_1 +\bar \eps_2) \psi_1 - {1\o 2} \eps_- \psi_1 \tilde \psi_1 + 
 \bar \eps_+ \( \psi_2e^{-{1\o 2} R_1} + {1\o 2} \psi_1 \tilde \chi_1\) , \nonu \\
 \d \chi_1 &=& {1\o 2} (\eps_1 +\eps_2 - \bar \eps_1 -\bar \eps_2) \chi_1 - {1\o 2} \bar \eps_+ \chi_1 \tilde \chi_1 +  
 \eps_- \( \chi_2e^{-{1\o 2} R_1} + {1\o 2} \chi_1 \tilde \psi_1\) , \nonu \\
 \d \psi_2 &=& \eps_- \({1\o 2} \psi_2 \tilde \psi_1 - \psi_1 e^{-{1\o 2} R_1} \) - {1\o 2} \bar \eps_+ \psi_2 \tilde \chi_1 + 
 {1\o 2} (-\eps_2 + \bar \eps_2)\psi_2, \nonu \\
 \d \chi_2 &=& \bar \eps_+ \({1\o 2} \chi_2 \tilde \chi_1 - \chi_1 e^{-{1\o 2} R_1} \) - {1\o 2}  \eps_- \chi_2 \tilde \psi_1 +
  {1\o 2} (\eps_2 - \bar \eps_2)\chi_2,
 \label{tranf44}
 \er
where $\eps_1 - \bar \eps_1, \eps_2 - \bar \eps_2, \eps_- $ and $\bar \eps_+$ are arbitrary constants.  The algebra of such transformations
can be shown to be the $q$-deformed Poisson bracket algebra $SL(2)_q \otimes U(1)$ \cite{inprep}, with $q= \exp (-{{2\pi }\o {k}})$.
 The global 
symmetries of the vector model generate the same algebra.

\section{Non-Conformal T-Duality}

T-duality in the context of the conformal $\s$-models 
\br
S_{\s}^{conf} = {{1\o {4\pi \a^{\pr}}}} \int d^2z \( (g_{MN}(X)\eta^{\mu \nu} + \eps^{\mu \nu} b_{MN}(X))\pa_{\mu} X^{M} \pa_{\nu} X^{N}
+ {{\a^{\pr}}\o 2} R^{(2)} \varphi(X) \) 
\label{act}
\er
($\mu , \nu = 0,1, \;\; M,N = 1,2, \cdots D$ and $R^{(2)} $ is the worldsheet curvature), represents specific canonical transformations (CT): 
$\(\Pi_{X_M}, X^M\) \rightarrow \(\Pi_{\tilde X_M}, \tilde X^M\) $ that map (\ref{act}) into its dual $\s$-model $S_{\s}^{conf}
\( G_{M,N}(\tilde X), B_{M,N}(\tilde X),\phi(\tilde X) \)$.  In the case of curved backgrounds with d-isometric directions (i.e., the metric
$g_{MN}(X^m)$, the antisymmetric tensor $b_{MN}(X^m)$ and the dilaton $\varphi (X^m)$ are independent of the
 $d\leq D$ fields $X_{\a}(z, \bar z), \a = 1, 2, \cdots d$) the corresponding CT has the form:
\br
 \Pi_{\tilde X_{\a}} = -2 \pa_x X_{\a}, \quad \Pi_{ X_{\a}} = -2 \pa_x \tilde X_{\a}
\label{cts}
\er
 and the others $\Pi_{ X_{m}}$ and $X_{m}, \; m=d+1, \cdots D$ remain unchanged. 
  Then T-duality manifests as (matrix) transformations of the target-space
 geometry data of (\ref{act}):
 $e_{MN}(X) = b_{MN}(X) + g_{MN}(X)$ and $\varphi(X) $ to its T-dual $E_{MN}(\tilde X) = B_{MN}(\tilde X) + G_{MN}(\tilde X)$ and
 $\phi(\tilde X)$ \cite{buscher}:
\br
E_{\a \b} = (e^{-1})_{\a \b} , \quad \quad E_{m n} = e_{m n} - e_{m \a} (e^{-1})^{\a \b} e_{\b n} \nonu \\
E_{\a m} = (e^{-1})_{\a}^{\b} e_{\b m}, \quad E_{m \a} =- e_{m \b }(e^{-1})^{\b}_{\a}, \quad \phi= \varphi - ln (det e_{\a \b})
\label{tr}
\er
By construction the dual pair of $\s$-models $S_{\s}^{conf}(e, \varphi )$ and $\tilde {S}_{\s}^{conf}(E, \phi )$ share the same spectra and
partition functions.  Their Lagrangians are related by the generating function ${\cal F}$ \cite{alvarez}
\br
{\cal L}(e, \varphi ) = {\cal L}(E, \phi ) + {{d{\cal F} }\o {dt }},  \quad  {\cal F} = 
{{1}\o {8\pi \a^{\pr}}}\int dx \(  X\cdot \pa_x \tilde X - \pa_x X \cdot \tilde X\) 
\label{func}
\er
An important feature of the abelian T-duality (\ref{cts}) and (\ref{tr}) is that it maps the $U(1)^{\otimes d}$ Noether charges $Q^{\a} =
\int_{-\infty}^{\infty} J_o^{\a} dx $ of $S_{\s}^{conf} (e, \varphi )$ into the topological charges 
$\tilde Q_{top}^{\a} = \int_{-\infty}^{\infty} \pa_x \tilde X^{\a} dx $ of its T-dual model $\tilde S_{\s}^{conf}(E, \phi)$ and vice-versa,
i.e., we have
\br
J_{\mu}^{\a} &=& e^{\a \b}(X_n) \pa_{\mu} X_{\b} + e^{\a m}(X_n) \pa_{\mu} X_m = \eps_{\mu \nu} \pa^{\nu} \tilde X^{\a}, \nonu \\
\tilde J_{\mu}^{\a} &=& E^{\a \b}(\tilde X_n) \pa_{\mu} \tilde X_{\b} + E^{\a m}(\tilde X_n) \pa_{\mu} \tilde X_m 
= \eps_{\mu \nu} \pa^{\nu}  X^{\a}
\label{curr}
\er
and therefore
\br 
T: (Q^{\a}, Q^{\a}_{top}) \rightarrow (\tilde Q^{\a}_{top}, \tilde Q^{\a})
\nonu
\er
Different examples of such T-dual pairs of conformal $\s$-models have been constructed in terms of axial and vector gauged $G/H$-WZW models
(see \cite{tseytlin}  and references therein).

From the other side the IM's considered in Sect. 2 and 3 have as their conformal limits ($\mu=0$, i.e. $V=0$ in (\ref{6.6}) and 
(\ref{svec})) the
corresponding axial and  vector gauged  $SL(3,R)/SL(2,R)\otimes U(1)$-WZW models which are T-dual by construction.  
They have $d=2$ isometric
directions, i.e., $e_{MN}(\psi_i, \chi_i)$ are independent of $\Theta_i = ln ({{\psi_i}\o {\chi_i}})$. 
 T-duality group in this case is known
to be $O(2,2|Z)$ (see for instance \cite{giveon}).  The problem we addresss in this section is about T-duality 
of the IM's (\ref{6.6})  and (\ref{svec}).  We first note the important property of these IM's, namely adding 
the potentials $V= Tr (\eps_+g_0^f \eps_- (g_0^{f})^{-1})$ breaks the conformal symmetry, but one still keep 
two isometries, i.e., $U(1)\otimes U(1)$ 
invariance, say
$\Theta_i \rightarrow \Theta_i +\a_i$ in the axial case.  This suggests that the T-duality of the conformal $G/H$-WZW models 
can be extended
to T-duality for their integrable perturbations (\ref{6.6}) and (\ref{svec}).  In order to prove it we extend the Buscher 
 procedure \cite{buscher}of
deriving T-dual of a given conformal $\s$-model (with $d$ isometries) to the case of IM's, i.e., in the presence of the potential $V(X_n)$.

\subsection{Isometries and T-Dual Actions}

Let us consider  the Lagrangian density of the form
\br
{\cal L}_{IM}^{ax} =  {\cal L}_{\s}^{conf}( \Theta_{\a}, X_m) - V(X_m)
\label{L}
\er
where ${\cal L}_{\s}^{conf}$  is the Lagrangian (\ref{act}) with $X_{\a} = \Theta_{\a}$ and the potential $V(X_m) $ is
 independent of $\Theta_{\a}$.  We next rewrite (\ref{act}) in a symbolic form separating the isometric fields $\Theta_{\a}, \; \a = 1,2,
 \cdots d$ from the remaining ones $X_m, \;\; m =d+1, \cdots D$:
 \br
{\cal L}_{IM}^{ax} =   \bar \pa \Theta_{\a} e^{\a \b} (X_m) \pa \Theta_{\b} +\bar \pa \Theta_{\a} N_{\a} + \bar N_{\a}\pa \Theta_{\a} 
+  {\cal L ^{\pr}}(X_m)
\label{Lax}
\er
In order to derive ${\cal L}_{IM}^{vec} (\tilde \Theta_{\a}, \tilde X_m )$ of the T-dual IM we apply eq. (\ref{func}), i.e.,
\br
{\cal L}_{IM}^{vec} ( \tilde \Theta_{\a}, \tilde X_m)= {\cal L}_{IM}^{ax} (  \Theta_{\a},  X_m) -\tilde \Theta_{\a}
\(\pa \bar P_{\a} - \bar \pa P_{\a} \) 
\label{lmodint}
\er
where we denote $P_{\a} = \pa \Theta_{\a}$, $\bar P_{\a} = \bar \pa \Theta_{\a}$ and the second term is nothing but the contribution of the
generating function ${\cal F}(\Theta_{\a}, \tilde \Theta_{\a}) \sim  \eps^{\mu \nu } \pa_{\mu} \Theta_{\a} \pa_{\nu}\tilde \Theta^{\a}$.  We
first integrate (\ref{lmodint}) by parts 
\br
{\cal L}_{IM}^{vec} = \bar P_{\a} e_{\a \b} P_{\b}  + \bar P_{\a} \( N_{\a} + \pa \tilde \Theta_{\a}\) 
+\(\bar N_{\a} - \bar \pa \tilde \Theta_{\a}\) P_{\a}   + {\cal L ^{\pr}}(X_m)
\label{lmodint2}
\er
and next we can take the Gaussian integral in $\bar P_{a}$ and $P_{\a}$ in the corresponding path integral.  Therefore the effective action
for the T-dual model has the form
\br
{\cal L}_{IM}^{vec} (\tilde \Theta_{\a}, X_m)=  
-\(\bar N_{\a} - \bar \pa \tilde \Theta_{\a}\)e^{-1}_{\a \b} \( N_{\b} + \pa \tilde \Theta_{\b}\) + {\cal L ^{\pr}}(X_m) - 4\pi (\a^{\pr})^2
ln (det e_{\a \b})R^{(2)}
\label{leff}
\er
in accordance with eqs. (\ref{tr}).

The second question to addressed is whether the Lagrangians (\ref{L}) and (\ref{leff}) are related by canonical transformations (\ref{cts}). 
In order to answer it, we shall compare their Hamiltonians:
\br
{\cal H}^{ax} = \dot {\Theta}_{\a} \Pi_{\Theta_{\a}} + \dot {X}_{m} \Pi_{X_{m}}- {\cal L}^{ax}, \quad  \quad 
{\cal H}^{vec} = \dot {\tilde {\Theta}}_{\a} \Pi_{\tilde \Theta_{\a}} + \dot {X}_{m} \Pi_{X_{m}}- {\cal L}^{vec} \nonu
\er
since by definition 
\br
\Pi_{\Theta_{\a}} &=& {{{\d {\cal L}^{ax}}\o {\d \dot {\Theta}_{\a}}}} = 2 \dot {\Theta_{\b}}e_{\a \b} + N_{\a} + \bar N_{\a}, \nonu \\
\Pi_{\tilde \Theta_{\a}} &=& {{{\d {\cal L}^{vec}}\o {\d \dot {\tilde \Theta}}_{\a}}} = 
e_{\a \b}^{-1}(2 \dot {\tilde \Theta_{\b}} + N_{\b} - \bar N_{\b}), 
\label{pi}
\er
we find that 
\br 
{\cal H}^{ax} &=&  {1\o 4} \Pi_{\Theta_{\a}}e_{\a \b}^{-1}\Pi_{\Theta_{\b}} - {1\o 2}\Pi_{\Theta_{\a}}e_{\a \b}^{-1}(N_{\b}  +\bar N_{\b}) 
+  \pa_x \Theta_{\a} e_{\a \b}\pa_x \Theta_{\b} + \pa_x \Theta_{\a} (N_{\a} - \bar N_{\a}) \nonu \\
&+&{1\o 4} (N_i + \bar N_i) e_{ij}^{-1}(N_j + \bar N_j) + {\cal H}(X_m, \Pi_{X_m})
\label{H}
\er
and 
\br 
{\cal H}^{vec} &=&  {1\o 4} \Pi_{\tilde \Theta_{\a}}e_{\a \b}\Pi_{\tilde \Theta_{\b}} - 
{1\o 2}\Pi_{\tilde \Theta_{\a}}(N_{\a}  -\bar N_{\a}) 
+  \pa_x \tilde \Theta_{\a} e_{\a \b}^{-1}\pa_x \tilde \Theta_{\b} + \pa_x \tilde \Theta_{\a} e_{\a \b}^{-1}(N_{\b} + \bar N_{\b}) \nonu \\
&+&{1\o 4} (N_i + \bar N_i) e_{ij}^{-1}(N_j + \bar N_j) + {\cal H}(X_m, \Pi_{X_m})
\label{Heff}
\er
where $ {\cal H}(X_m, \Pi_{X_m}) = \dot {X}_m \Pi_{X_m} - {\cal L}^{\pr} (X_m)$. 
Finally we observe that ${\cal H}^{ax} = {\cal H}^{vec}$, i.e. integrable models (\ref{L}) and (\ref{leff}) have coinciding 
Hamiltonians if the transformation  
\br
 \Pi_{\Theta_{\a}} = -2\pa_x \tilde \Theta_{\a}, \quad \quad  \Pi_{\tilde \Theta_{\a}} = -2\pa_x \Theta_{\a}
\label{cc}
\er
takes place.  This is precisely the canonical transformation (\ref{cts}) relating the T-dual pais of $\s$-models.

\subsection{Axial-Vector Duality for Homogeneous Grading Models}
In order to prove that the axial (\ref{6.6})  and vector (\ref{svec}) IM's are T-dual to each other, we apply the procedure explained in
Sect. 5.1.  Starting  from eq. (\ref{6.6}) we recognize the two isometric ``coordinates '' to be  $\Theta_{\a} = ln ({{\psi_{\a}}\o
{\chi_{\a}}}), \; \a =1,2$.  By changing variables 
\br
\psi_{\a}, \chi_{\a} \rightarrow \Theta_{\a}, \; a_m = \psi_m \chi_m, \; m=1,2
\nonu 
\er
one can rewrite ${\cal L}^{ax}$  in (\ref{6.6}) in the form (\ref{Lax}) with
\br
{\cal L}^{\pr}(X_m) &=&   {{\bar \pa a_1 \pa a_1}\o {4\Delta a_1}}(1+a_2) +
  {{\bar \pa a_2 \pa a_2}\o {4\Delta a_2}}(1+a_1+a_2) -  {{\bar \pa a_1 \pa a_2}\o {8\Delta }} 
-  {{\bar \pa a_2 \pa a_1}\o {8\Delta }} -  {{\mu^2}}({2 \o 3} +a_1 +a_2) \nonu 
\label{ref}
\er
and 
\br
e_{11} = -{1\o {4\Delta}} (1+a_2) a_1, \quad e_{22} = -{1\o {4\Delta}} (1+a_1+a_2) a_2,\quad e_{12} = e_{21} = {{1\o {8\Delta}}}a_1a_2
\nonu
\er
\br
N_1 &=& {{1\o {4\Delta}}}((1+a_2)\pa a_1 - {1\o 2}a_1\pa a_2), \quad N_2 ={{1\o {4\Delta}}}((1+a_1+a_2)\pa a_2 - {1\o 2}a_2\pa a_1), \nonu \\
\bar N_1 &=& {{1\o {4\Delta}}}(-(1+a_2)\bar \pa a_1 + {1\o 2}a_1\bar \pa a_2), \quad 
\bar N_2 =-{{1\o {4\Delta}}}((1+a_1+a_2)\bar \pa a_2 - {1\o 2}a_2\bar \pa a_1)
\label{mnn1}
\er
Therefore, according to eqns. (\ref{lmodint}) and (\ref{lmodint2}) the axial and vector IM's are 
related by canonical transformation (\ref{cc}).  The
identification of ${\cal L}_{IM}^{vec}$  in (\ref{leff}) with the vector model Lagrangian 
(\ref{svec}) becomes evident by observing
the relations among the fields,
\br
a_1 = -t_1^2 e^{\phi_1 + \phi_2}, \quad \quad a_2 =-t_2^2 e^{-\phi_1 +2\phi_2}, \quad  \quad 
\tilde \Theta_1 = -{1\o 2} \phi_1, \quad \quad \tilde \Theta_2 = -{1\o 2} (\phi_2 -\phi_1).
\label{5.15}
\er

Another important feature of the axial-vector  T-duality is the simple relation between the isometric filds $\tilde \Theta_{\a}$ of the
vector model (\ref{svec}) and the nonlocal fields $R_i$ (see (\ref{rr}) ) of the axial model,
\br
R_1 = 2 (\tilde \Theta_2 - \tilde \Theta_1 ), \quad \quad R_2 = -2 (\tilde \Theta_1 +2\tilde \Theta_2 )
\label{5.16}
\er
The above identification can be established by solving the constraints (\ref{2.7b})  (or in the explicit 
form (\ref{rr}) for the $SL(3)$ case)  in favour of the non local fiels of the vector model  $\Theta_i$:
\br
\pa \Theta_1 &=& \pa ln a_1 - \pa (R_1 +R_2) - {2\o 3} {{a_2+1}\o {a_1}} \pa (2R_1 + R_2), \nonu \\
\pa \Theta_2 &=& \pa ln a_2 + {2\o 3} {{a_2+1}\o {a_2}}  \pa (R_1 -R_2) - {1\o 3} \pa (2R_1 +R_2), \nonu \\
\bar \pa \Theta_1 &=& -\bar \pa ln a_1 +\bar \pa (R_1 +R_2) + {2\o 3} {{a_2+1}\o {a_1}} \bar \pa (2R_1 + R_2), \nonu \\
\bar \pa \Theta_2 &=& -\bar \pa ln a_2 - {2\o 3} {{a_2+1}\o {a_2}}  \bar \pa (R_1 -R_2) + {1\o 3} \bar \pa (2R_1 +R_2)
\label{5.17}
\er
 and next comparing the RHS of eqn. (\ref{5.17})  with the $U(1)\otimes U(1)$ conserved currents of the 
 vector model Lagrangian (\ref{svec}).  We can further write eqns. (\ref{5.17}) 
 and (\ref{rr}) in the compact form
 \br
 J_{top}^{i, ax} = \eps_{\mu \nu } \pa^{\nu} \Theta_i = \tilde J_{\mu}^{i, vec}, \quad
 \tilde J_{top}^{i, vec} = \eps_{\mu \nu } \pa^{\nu} R_i =  J_{\mu}^{i, ax}, 
 \label{5.18}
\er
or equivalently
\br
 \tilde I_{top}^{1, vec} &=& \eps_{\mu \nu } \pa^{\nu} \tilde \Theta_1 = -{1\o 6} (J_{\mu}^{2, ax} + 2 J_{\mu}^{1, ax}), \nonu \\
 \tilde I_{top}^{2, vec} &=& \eps_{\mu \nu } \pa^{\nu} \tilde \Theta_2 =  {1\o 6} (J_{\mu}^{1, ax} - J_{\mu}^{2, ax}), 
 \label{5.19}
\er
These equations examplify for the $SL(3)$-case in consideration the main property (\ref{curr}) of the T-dual pairs of models 
\br
Q_{top}^{\a, ax} = Q^{\a, vec}, \quad \quad Q_{top}^{\a, vec} = Q^{\a, ax}
\label{5.20}
\er
namely that T-duality relates the topological charges $Q_{top}^{\a, vec} = \int dx \pa_{x} \tilde \Theta_{\a}$ to the 
$U(1)\otimes U(1)$-charges $Q^{\a, ax}$ of the axial IM and vice-versa.

An explicit realization of the above exchange of topological and $U(1)$-Noether charges (similar to the momentum-winding 
numbers exchange in
string theory) have been observed in ref. \cite{elek}, analyzing the 1-soliton structure spectrum of the corresponding 
dyonic IM.  The masses
of the solitons of axial and vector models remains equal, but the $U(1)$ charge of the axial non-topological solitons 
is transformed into the topological charge of the vector model solitons.  Similar relations take place in the pair of T-dual nonabelian
dyonic models (\ref{6.6}) and (\ref{svec}) in consideration \cite{inprep}.

\section{Conclusions}
We have demonstrated how one can extend the abelian T-duality of the conformal gauged $G/H$-WZW models to their integrable perturbations,
that appears to be identical to specific homogeneous gradation NA affine Toda models.   More general
 considerations (presented in Sect. 5) of
generic (relativistic) IM's (as well as for non integrable models) 
admiting isometric directions (i.e., with few global $U(1)$ symmetries) make evident 
that one can construct their
T-dual partners by approprietly chosen cannonical transformations.  The most important new feature of the 
T-duality in the context of 2-D
integrable models consists in its action  on the spectrum of the solitons of the corresponding pair of dual IM's. 
 As one can expect it maps the
$U(1)^{\otimes d}$-charges of the solitons of the axial model (with $d$-isometries)  to the topological 
charges of the solitons of its T-dual
counterpart, leaving the soliton masses unchanged.

The quantization of the NA affine Toda models usualy require nontrivial counterterms \cite{elek}, \cite{Fat},
 \cite{devega} together with the
renormalization of the couplings and masses.  Hence, an interesting open problem is whether the quantum vector 
and axial IM's continue to be
T-dual to each other.

{\bf Acknowledgments}  One of us (JFG) thanks O. Babelon,  for discussions and LPTHE for the hospitality.
We are grateful to  CNPq, FAPESP,  UNESP and CAPES/COFECUB for 
financial support.

\end{document}